\documentclass[10pt]{article}
\usepackage[OE]{express}
\usepackage{cite}
\begin{document}

\title{Serial--parallel conversion for single photons with heralding signals}

\author{Takayuki Kiyohara$^1$, Ryo Okamoto$^{1,2}$, and Shigeki Takeuchi$^{1,*}$}

\address{$^1$Department of Electronic Science and Engineering, Kyoto University, Kyoto Daigaku-Katsura, Nishikyo-ku, Kyoto 615-8510, Japan.\\
$^2$PRESTO, Japan Science and Technology Agency, 4-1-8 Honcho, Kawaguchi, Saitama 332-0012, Japan}

\email{*takeuchi@kuee.kyoto-u.ac.jp} %% email address is required



\begin{abstract}
We present serial--parallel conversion for a heralded single photon source (heralded SPS). We theoretically show that with the heralding signal the serial--parallel converter can route a stream of $n$ photons to $n$ different spatial modes more efficiently than is the case without using a heralding signal. We also experimentally demonstrate serial--parallel conversion for two photons generated from a heralded SPS. We achieve a conversion efficiency of $0.533\pm0.003$, which exceeds the maximum achievable efficiency of 0.5 for serial--parallel conversion using unheralded photons, and is double the efficiency (0.25) for that using beamsplitters. The efficiency of the current setup can be increased up to $0.996\pm0.006$ when the losses in the optical converter are corrected for.
\end{abstract}

 % REPLACE WITH CORRECT OCIS CODES FOR YOUR ARTICLE, MINIMUM OF TWO; Avoid using the OCIS codes for "General???or "General science???whenever possible.
%For a complete list of OCIS codes, visit: https://www.osapublishing.org/oe/submit/ocis/

%%%%%%%%%%%%%%%%%%%%%%% References %%%%%%%%%%%%%%%%%%%%%%%%%

%%%%%%%%%%%%%%%%%%%%%%%%%%  body  %%%%%%%%%%%%%%%%%%%%%%%%%%
\section{Introduction}
To realize large scale photonic quantum technologies, it is important to use a source that emits multiple indistinguishable single photons \cite{Knill,Kimble,Pan,Scarani,Aspuru}. Until now, heralded single photon sources (heralded SPS) using spontaneous parametric down-conversion (SPDC) have been widely used because of the high indistinguishability of the photons generated by these sources \cite{Tanida}. An alternative scheme, multiplexed heralded SPS, has been proposed and demonstrated, which can maintain the probability of generating one photon while suppressing the probability of having more than two photons \cite{Ma,Collons,Mendoza,Kaneda,Xiong,Kiyohara}. However, this scheme requires considerable resource overhead in terms of multiple optical switches and heralded SPSs, since a multiplexed heralded SPS requires multiple nonlinear crystals, detectors and other elements. A further alternative method comprising a single photon source using single emitters such as quantum dots (QD) has been developed recently \cite{Somaschi,Ding}, although achieving multiple and identical single emitters is still a challenge.

To address this difficulties in multiple single photon sources, a serial--parallel conversion method has been proposed \cite{QIT,JPS}. The converter can route a stream of $n$ photons at successive intervals from one SPS to $n$ different spatial modes. Until now, a converter using a passive router consisting of a balanced beam splitter has been widely used \cite{Loredo}. Recently, Lenzini et al. \cite{Lenzini} and Wang et al. \cite{Wang} have realized active serial--parallel conversion for single photons generated from a quantum dot, where photons were produced without heralding signals.

In this paper, we report the experimental realization of a serial--parallel conversion system for two photons with heralding signals. We theoretically compared the conversion efficiency for $n$ heralded photons with that for $n$ unheralded photons and found that serial--parallel conversion for the heralded photon source could be more efficient than for the unheralded photon source. With the realized two-photon serial--parallel converter, we confirmed the efficacy of serial--parallel conversion for the case where a series of two heralding signals is detected. We achieved a conversion efficiency of $0.533\pm0.003$, which exceeds the maximum achievable efficiency of 0.5 for serial--parallel conversion using unheralded photons, and is double the efficiency of 0.25  for that using beamsplitters. The efficiency of current setup can be increased up to $0.996\pm0.006$ when the losses in the optical converter are corrected for.

\section{Model of serial--parallel converter}
Figure 1 shows an example of the scheme for a serial--parallel converter, where four photons are converted into four different spatial modes. A nonlinear crystal is pumped by a pulsed laser and generates photon pairs via the SPDC process. When one of the photons in the pair is detected, a heralding signal is generated. Using the heralding signals, the controller identifies which pulses have corresponding photons, and generates an $n$-photon heralding signal when a stream of $n$ photons is detected. In this example, a series of four photons from a heralded SPS $n$ = 4 is routed into the four different spatial modes by the serial--parallel converter consisting of three optical routers and delay lines. Note that for $n$-mode output, the number of required routers is $n-1$. The converted photons are output to the four modes at the same time using optical delay lines.

\begin{figure}[t] \centering\includegraphics[width=10cm]{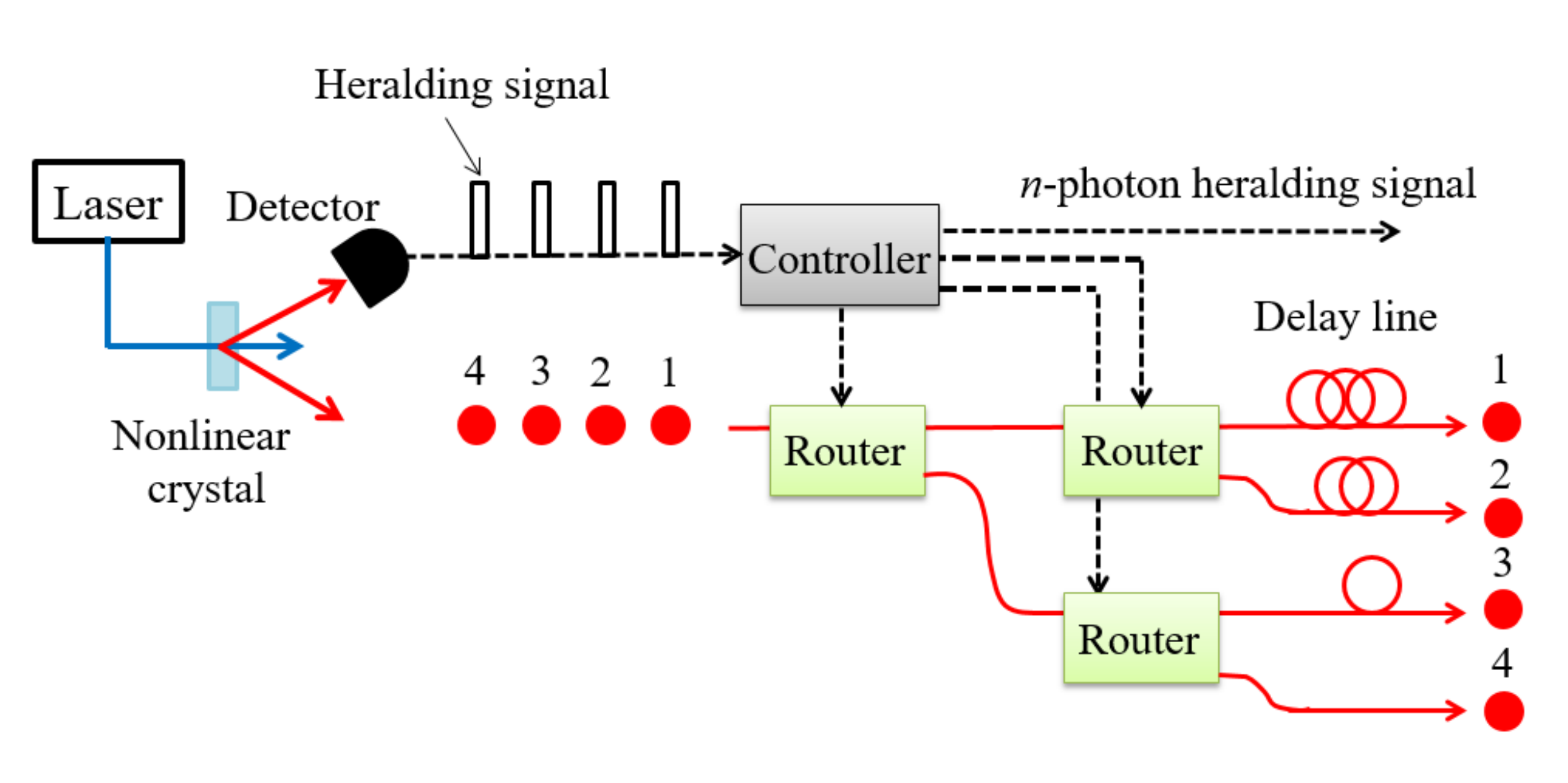}
\caption{ Scheme of serial--parallel conversion for four photons generated from a heralded single photon source $n$ = 4. A nonlinear crystal is pumped by a pulsed laser and generates photon pairs via the SPDC process. A stream of photons (Red dots) is intendedly routed into different spatial modes by a convertor consisting of the routers and delay lines. The controller generates the $n$-photon heralding signal and the signal of controlling the routers when a stream of $n$ photons is detected at the heralding arm. Red lines connecting the routers represent the path of photons. The dotted arrows from the controller to the router represent the path of electrical signal.}
\end{figure}

\section{Comparison of conversion efficiencies between heralded and unheralded single photons}
Here, we compare the conversion efficiency $S (n)$ for a heralded SPS with that for an unheralded SPS. $S (n)$ is a parameter that represents how efficiently an arbitrary $n$-photon stream is converted into different spatial modes using a serial--parallel converter. 

  In the case of heralded photons, we consider the conversion efficiency $S(n)$ for the case where a series of $n$ photon heralding signals is detected. Then, the optical routers are driven using the heralding signals, which indicate the presence of the corresponding photons. When the heralded SPS emits a stream of $n$ photons, the $n$-photon heralding signal identifies the presence of the generated $n$ photons precisely. For example, figure 2(a) shows that a heralded SPS emits a stream of four photons, and the corresponding heralding signals. The ideal converter perfectly transfers a stream of heralded single photons to different optical modes. For the case where a $n$-photon heralding signal is generated, the conversion efficiency for a stream of $n$ photons is given by
\begin{equation}
S (n)=\eta _{\rm SW}^n,
\end{equation}
where $\eta_{\rm SW}$ is the switching efficiency defined as the average probability of routing a single photon in the desired modes. 

Next, let us consider the case of single photons without heralding signals \cite{Lenzini} [Fig.2(b)]. In this case, the serial--parallel converter routes the photons automatically according to the clock signals provided by the source (e.g., a pumping laser) [see Fig.2(b)]. However, in this case, the routing of a stream of four photons is only successful when the `1st clock signal' by chance coincides with the existence of the first photon among a stream of four photons.  For this case, the conversion efficiency is given by

\begin{equation} 
S(n)=\frac{1}{n}\Bigl[\eta _{\rm SW}^n + (n-1)\Bigl(\frac{1-\eta_{\rm SW}}{n-1}\Bigr)^n \Bigr],
\end{equation}

\begin{figure}[t] \centering\includegraphics[width=9.5cm]{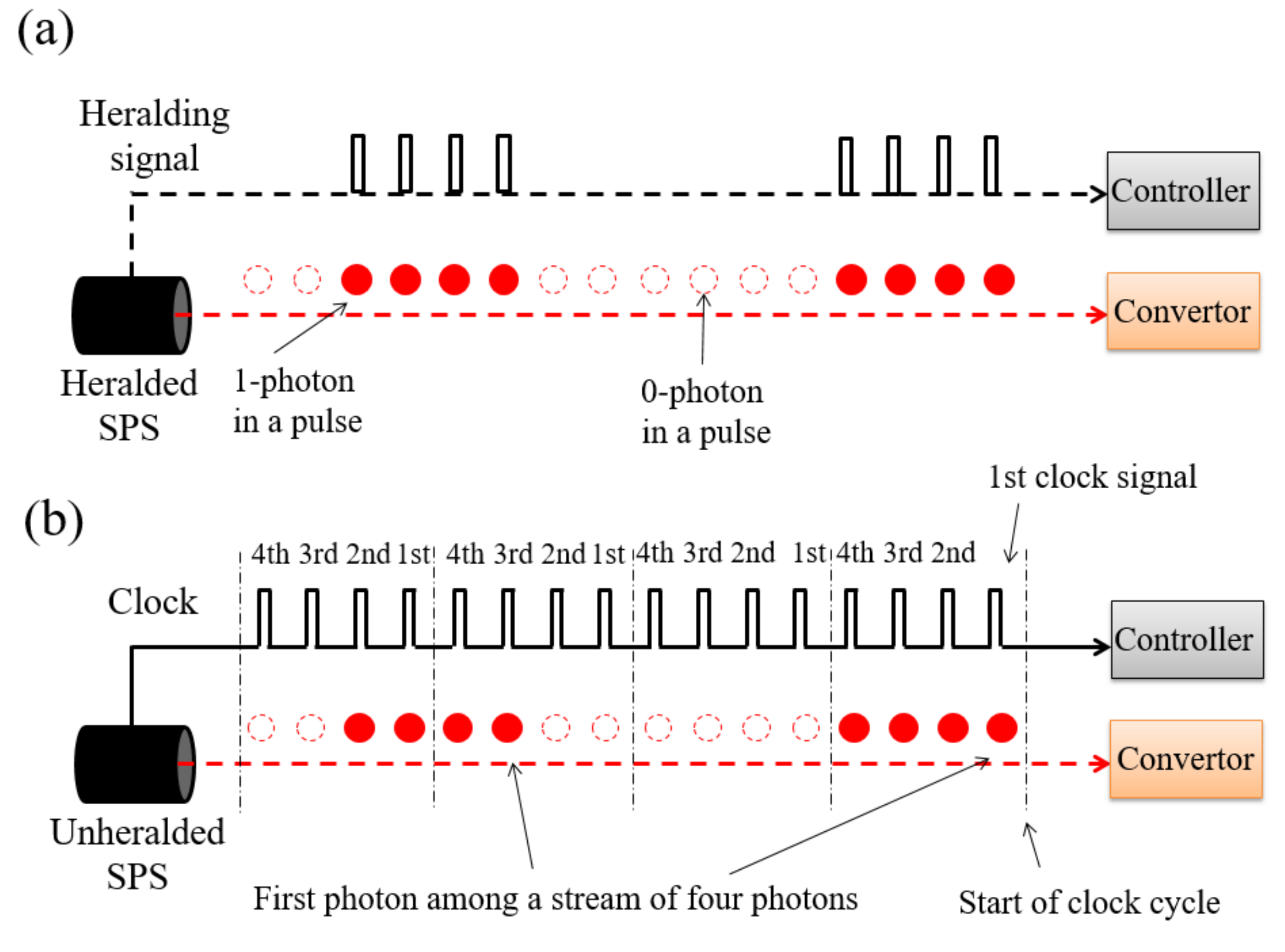}
\caption{(a) Heralded SPS and serial--parallel converter. A heralded SPS emits a stream of four photons and the corresponding heralding signals. (b) Unheralded SPS and serial--parallel converter. An unheralded SPS emits a stream of four photons and the clock signal of the pump pulse laser. The clocking is derived from the pump laser. The converter operates according to the interval of a four-clock cycle. The conversion succeeds only when the first clock from the divided line corresponds to the first photon among the four photons.}
\end{figure}

\begin{figure}[t] \centering\includegraphics[width=9cm]{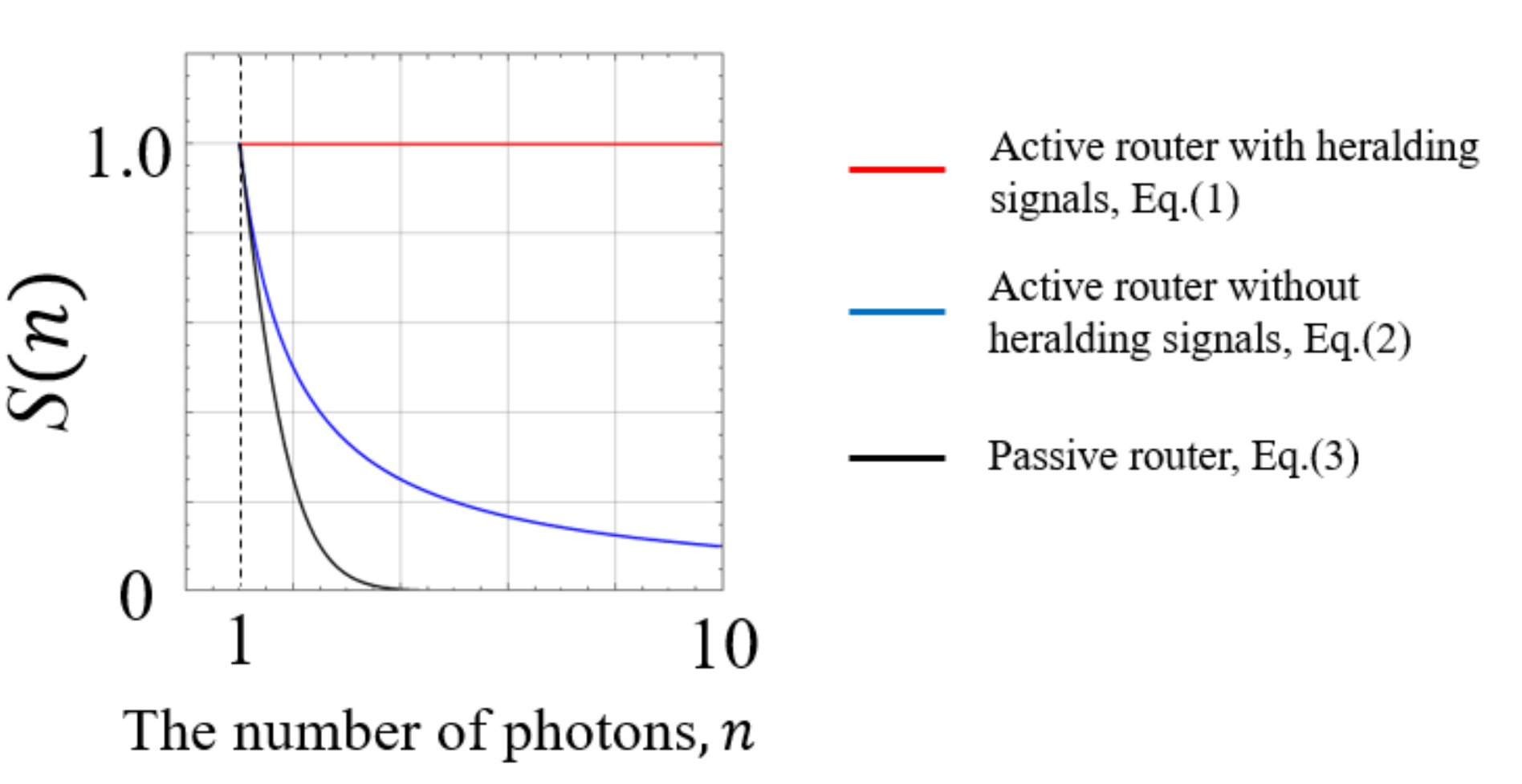}
\caption{Theoretical comparison of the efficiencies of serial--parallel conversion using active routers with heralding signals (red line), active routers without heralding signals (blue line) and passive routers (black line). Here an ideal switching efficiency is assumed.  }
\end{figure}
 
Finally, if a beam splitter is used as a passive router, \cite{Loredo}, the efficiency for heralded photons is at most
\begin{equation} 
 S(n) = (1/n)^n,
\end{equation} 
 which can only be attained when the loss of the beamsplitters is negligibly small.

Figure 3 shows the plots of eqs.(1) -- (3) for an ideal switching efficiency $\eta_{\rm SW}=1$. As $n$ is increased, conversion efficiencies $S(n)$ of the passive router (black line) and the active router without heralding signals (blue line) degrade, while $S(n)$ of the active router with heralding signals (red line) is maintained. Comparing these conversion efficiencies shows that serial--parallel conversion using active routers with heralding signals is more efficient than those using active routers without heralding signals, or passive routers obviously.

\section{Experimental setup of serial--parallel conversion for two photons}
\begin{figure}[h] \centering\includegraphics[width=12cm]{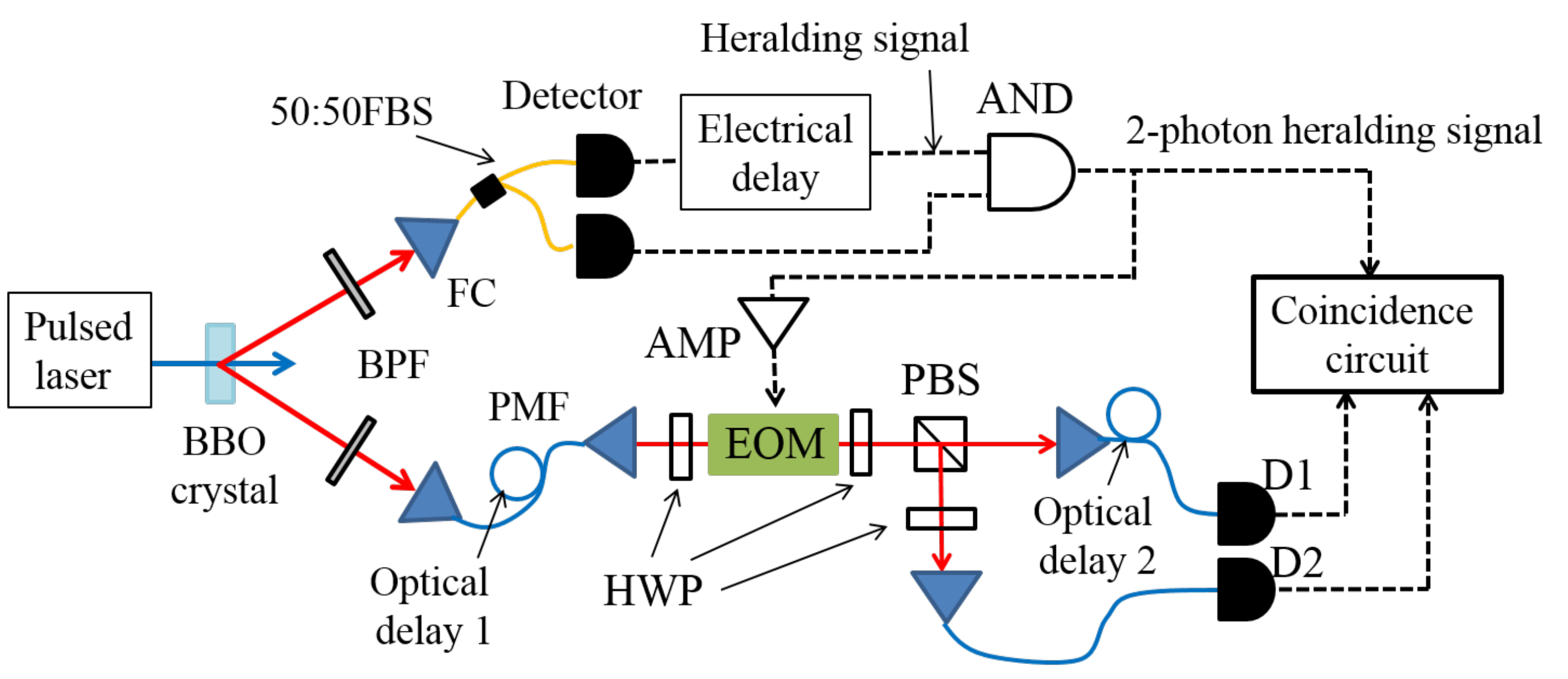}
\caption{Schematic diagram of experimental setup.  BPF: band pass filter, HWP: half wave plate, FC: fiber coupler, 50:50FBS: 50:50 fiber beam splitter, AND: AND gate, PMF: polarization maintaining fiber, AMP: high-voltage pulsed amplifier, PBS: polarizing beam splitter, EOM: electro optics modulator. Optical fiber delay 1 is 65 m long. Optical fiber dealy 2 is 2.5 m long. }
\end{figure}
 We implemented a serial--parallel converter for two photons with heralding signals. A schematic of the experimental setup is shown in Fig.\ 4. A femtosecond pulse laser (Tsunami with second harmonic generation, Spectra Physics, repetition rate of 82 MHz and central wavelength of 390 nm) is used to pump a beta barium borate (BBO) crystal (Type-I, thickness of 1.5 mm), and generates photon pairs (signal photons and idler photons). For the SPDC source, the idler photons are detected by two detectors (SPCM-AQR, Excelitas Technologies) because the detector's deadtime (40 ns) is larger than the interval (12.5 ns) between the pump pulses. Using the electrical delay line and an AND logic gate, we select the cases where a series of two heralding signals exist. Then the electric signal (final heralding signal) is sent to the coincidence circuit (SR 400, Stanford Research Systems) and an electro optics modulator (EOM, LM0202, LINOS). 

The signal photons are sent to an optical converter consisting of half wave plates (HWP) and the EOM. Optical delay 1 (65 m) is inserted to correct for the delay caused by the electric signal processing mentioned above. With an appropriate timing adjustment, the EOM rotates the polarization of the second photon just after the first signal photon passes the EOM. Then, the first and second signal photons are routed into different spatial modes after a polarizing beam splitter (PBS) and are detected by detectors D1 and D2. Optical delay 2 corresponds to the ``delay line'' in Fig. 1.

We estimate the conversion efficiency $S(n)$ of our converter based on the experimentally obtained parameters $P_{\rm h}(1)$, the probability of having one photon in a pulse when the heralding signal is generated, $C_{\rm h}(n)$, the rate of events where a series of $n$ heralding signals is generated, and $C(n)$, the rate of events where $n$ photons are successfully detected at the same time at each of $n$ detectors when the series of $n$ heralding signals are generated. The conversion efficiency is thus given by

\begin{equation}
S (n)=\frac{C(n)}{C_{\rm h}(n)}\frac{1}{[P_{\rm h}(1)\eta_{\rm D}]^n},
\end{equation}
where the non-unity efficiency of the source ($P_{\rm h}(1)$) and the photon detectors ($\eta_{\rm D}$) are corrected to evaluate the performance of the serial--parallel converter.

\section{Experimental results and discussion}

Using Eq. (4), we evaluated the conversion efficiency $S (2)$ for the case where a series of two heralding signals is detected, to check the performance of our serial--parallel converter. 

In our experiment, the rate of the measured coincidence events $C(2)$ between the final heralding signals and the two output signals from detectors D1 and D2 was $2.62\pm0.15$ cps, indicating that the photons are sent to both D1 and D2 simultaneously with this rate. Note that the coincidence rate was $0.006\pm0.003$ cps when EOM was not operated by the final heralding signal, indicating that both photons are sent to only one of the two detectors (D1) with these results. We confirmed the successful operation of the serial--parallel conversion. 

The rate of the $n$-photon heralding signal $C_{\rm h} (2)$ is $785\pm8$ cps, and the one-photon probability from HSPS, including the detection efficiency, $P_{\rm h} (1)\eta _{\rm D}$ is $0.079\pm0.001$. For these measured parameters, we evaluated the conversion efficiency, using Eq.(4), to be $S (2)=0.533\pm0.003$. To confirm the advantage of our serial--parallel converter, we compared the conversion efficiency using routers for unheralded photons with the ideal switching efficiency ($\eta_{\rm SW}$ = 1)  . For $n$ = 2, the optical converter would enable a maximum conversion efficiency of 0.5  [Table 1]. The conversion efficiency of our optical converter is 0.533, which exceeds the maximum achievable efficiency for serial--parallel conversion using unheralded photons, and is double the efficiency (0.25) for that using beamsplitters.

\begin{table}[b]
  \caption{Theoretical conversion efficiencies $S(n)$ for an ideal switching efficiency ($\eta_{\rm SW}=1$), using Eqs.(1)-(3): an active router with heralding signals, an active router without heralding signals, a passive router using balanced beam splitters.}
  \label{table:data_type}
  \centering
   {\renewcommand\arraystretch{1.5}
  \begin{tabular}{|c|c|c|c|}
\hline
& Active router  &   Active router  &  Passive router        \\
& with heralding signals  &   without heralding signals &    \\
\hline

$S(n)$&1&1/$n$&$(1/n)^n$\\
\hline
$S(2)$&1&0.5&0.25\\
\hline
  \end{tabular}
  }
\end{table}

\begin{table}[b]
  \caption{Experimental losses and measured/compensated conversion efficiency in this work. }
  \label{table:data_losses}
  \centering
  {\renewcommand\arraystretch{1.5}
    \begin{tabular}{|c|c|c|}
 \hline
 Description & Symbol & Value in this work\\
 \hline
 \hline
  Observed conversion efficiency &$S(2)$ &0.533 $\pm$ 0.003\\
\hline
   Average transmittance of the converter & $t$ & 0.731 $\pm$ 0.003 \\
\hline
   Transmittance-compensated conversion efficiency &$S(2)/t^2$ &0.996 $\pm$ 0.006\\
\hline
  \end{tabular}
  }
\end{table}

The experimental losses and measured/compensated conversion efficiency are written in Table 2. The average transmittance ($t$) of the converter was 0.731 in the actual experimental setup. The corrected conversion efficiency is estimated to be $0.996\pm0.006$ when the optical loss in the converter ( between a single mode fiber (Optical delay 1) and the output ) is compensated. The corrected conversion efficiency of 0.996 is in good agreement with ideal value $S(2)$ =1 for active router with heralding signals [Table.1]. This result also confirms the success of our serial--parallel conversion for a series of two heralding signals.

To verify the successful operation of our router, we further measured the routing efficiency $\eta_{\rm 1}$ or $\eta_{\rm 2}$ which is defined as the probability of routing photon 1 or 2 into output 1 or 2, respectively, whereas the probability of routing photon 1 or 2 into output 2 or 1, respectively, is $1-\eta_{\rm 1}$ or $1-\eta_{\rm 2}$, as shown in	 Fig.\ 5(a). Figure\ 5(b) and 5(c) show the theoretical routing efficiencies using (b) a passive router ($\eta_{\rm 1} = \eta_{\rm 2} = 0.5$) and (c) an active router ($\eta_{\rm 1}  =\eta_{\rm 2} = 1$). In our experiment, the routing efficiency is estimated using coincidence events between the series of two heralding signals and the detection signal from detector D1 or D2, and found to be $\eta_{\rm 1}=0.99\pm0.01$ and $\eta_{\rm 2}=0.98\pm0.01$, shown in Fig.5(d). Note that the routing efficiency is corrected based on the optics components and detection efficiency. This measured value is in good agreement with the predicted value in Fig.\ 5(c). Thus, it can be said that our router can properly guide photons into the appropriate spatial modes. Note that, for our experimental setup, the switching efficiency $\eta_{\rm SW} = t \times (\eta_1 + \eta_2)/2=0.72$.

\begin{figure}[t]
\centering\includegraphics[width=12cm]{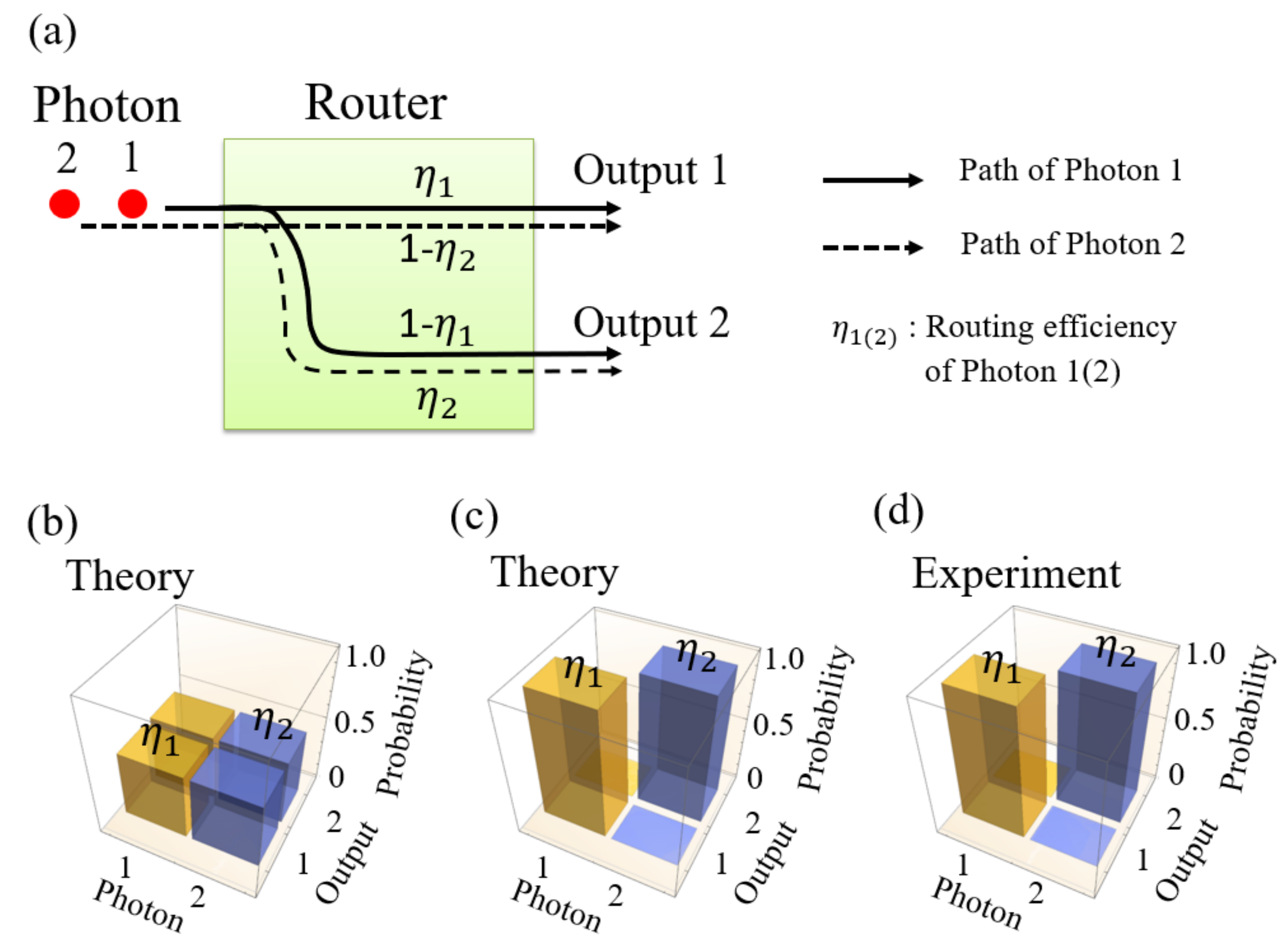}
\caption{(a) Sketch of an optical router. This routes photon 1 into output 1 with routing efficiency $\eta_{\rm 1}$, and routes photon 2 into output 2 with efficiency $\eta_{\rm 2}$. (b,c,d) Routing efficiency.  (b) Theoretical efficiency for a passive router using a balanced beam splitter. (c) Theoretical efficiency for an active router with heralding signals. (d) Experimental efficiency for an active router with heralding signals.}
\end{figure}

We note that our serial--parallel converter can be easily expanded for $n$ photon streams by adding EOMs and PBSs. For $n$ photons, $n-1$ EOMs will be required. The next challenge will be to generate a state of multiple photons in parallel. For such experiments, improving the heralded source with higher $P_{\rm h}(1)$ and suppressing excess photon emission is important. Recently, a $P_{\rm h}(1)$ of about 0.7 has been reported \cite{Wang2} using type-II beamlike conditions \cite{Takeuchi1,Zhang,Wang2}. Furthermore, the suppression of multi-photon component has been reported using various schemes \cite{Ma,Collons,Mendoza,Kaneda,Xiong,Kiyohara}.

\section{Conclusion}
In conclusion, we realized a serial--parallel conversion for two photons with heralding signals.  We theoretically compared the conversion efficiency for $n$ heralded photons with that for $n$ unheralded photons and found that a serial--parallel conversion using heralding signals could be more efficient.  With the realization of the two-photon serial--parallel converter, we confirmed the success of serial--parallel conversion for the case of two heralding signals. As a result, we achieved a conversion efficiency of $0.533\pm0.003$, which exceeds the maximum achievable efficiency of 0.5 for serial--parallel conversion using unheralded photons, and is double the efficiency (0.25) possible using beamsplitters. The efficiency of current setup can be increased up to $0.996\pm0.006$ when the losses in the optical converter are corrected for.

\section*{Funding}
JSPS-KAKENHI (No. 26220712, 26610052, 25707034), JST PRESTO (No. 10435951), JST-CREST, JSPS-FIRST, Project for Developing Innovation Systems of MEXT, Research Foundation for Opto-Science and Technology, The Research Fellow of Japan Society for the Promotion of Science (No. 17J06932).

\section*{Acknowledgments}

We wish to thank Prof. M. Shiraishi and Prof. T. Asano for helpful comments.

\end{document}